\documentclass[%
reprint,
preprintnumbers,
amsmath,amssymb,
aps,
]{revtex4-1}

\usepackage{graphicx}
\usepackage{dcolumn}
\usepackage{bm}
\usepackage{hyperref}
\usepackage{siunitx}
\usepackage{placeins}
\usepackage{makecell}
\usepackage{xcolor}
\definecolor{darkergreen}{rgb}{0,0.5,0}
\begin{document}
	
	\preprint{DO-TH 19/22}
	
	\title{Correlating uncertainties in global analyses within  SMEFT matters}
	\author{Stefan Bi\ss{}mann}
	\email{stefan.bissmann@tu-dortmund.de}
	\author{Johannes Erdmann}
	\email{johannes.erdmann@tu-dortmund.de}
	\author{Cornelius Grunwald}
	\email{cornelius.grunwald@tu-dortmund.de}
	\author{Gudrun Hiller}
	\email{ghiller@physik.uni-dortmund.de}
	\author{Kevin Kr\"oninger}
	\email{kevin.kroeninger@tu-dortmund.de}
	\affiliation{Fakult\"at  Physik, TU Dortmund, Otto-Hahn-Str.4, D-44221 Dortmund, Germany}

	\begin{abstract}
		We investigate the impact of correlations between (theoretical and experimental) uncertainties on multi-experiment, multi-observable analyses within the Standard Model Effective Field Theory (SMEFT). 
		To do so, we perform a model-independent analysis of $t$-channel single top-quark production and top-quark decay data from ATLAS, CMS, CDF and D0.
		We show quantitatively how the fit 
		changes when different  experimental or theoretical correlations are assumed.
		Scaling down statistical uncertainties according  to the luminosities of  future colliders
		with  $300 \, {\rm fb}^{-1}$ and higher, we find that this effect becomes a matter of life and death:
		assuming no correlations returns a fit in agreement with the Standard Model while a 'best guess'-ansatz taking into account correlations would observe new physics.
		At the same time, modelling the impact of higher order SMEFT-corrections the latter  turn out to be a subleading source of uncertainty only.
	\end{abstract}
	
	\maketitle

	\section{Introduction}
	\label{Sec:Introduction}
	The Standard Model Effective Field Theory (SMEFT) allows for model-independent analyses of multi-experiment, multi-observable data \cite{Weinberg,Buchmuller:1985jz,Grzadkowski:2010es}. 
	Higher-dimensional operators, built from Standard Model (SM) fields consistent with $SU(3)_C \times SU(2)_L\times U(1)_Y$- and Lorentz-invariance,
	systematically account for physics beyond the SM (BSM) at the scale $\Lambda$  above the electroweak one,  in an expansion in $1/\Lambda$.
	Along the lines of  high luminosity precision programs in flavor physics \cite{Cerri:2018ypt},
	new physics at scales beyond the colliders energy reach is probed indirectly, making global fits and uncertainty management key tasks.
	
	Correlations between measurements play an important role in global fits, e.g., recently discussed in \cite{Hartland:2019bjb,Durieux:2019rbz,Brivio:2019ius}. 
	However, the quantitative impact of such correlations continues to be unknown. In practice, this has led to  simplifying assumptions as well as the exclusion of data sets from  global fits.
	A first toy study of correlations has been discussed in Ref.~\cite{Bissmann:2019gfc}.
	
	In this paper, we work out how correlations of systematic uncertainties and theory uncertainties between measurements change fit results  within the SMEFT framework. 
	We entertain the example of $t$-channel single top-quark production together with top-quark decay; it is rather compact due to the small number of contributing Wilson coefficients, while still covers all relevant aspects of a global fit with 
	various observables from different experiments.
	In the recent past, several studies of the top-quark sector of SMEFT have been performed, see, for instance, Refs.~\cite{Degrande:2018fog,Chala:2018agk,Durieux:2014xla,AguilarSaavedra:2010zi,DHondt:2018cww,
		Durieux:2018ggn,Buckley:2015nca,Buckley:2015lku,deBeurs:2018pvs,Brown:2019pzx,AguilarSaavedra:2018nen,Hartland:2019bjb,Maltoni:2019aot,Durieux:2019rbz,Neumann:2019kvk,Brivio:2019ius,Bissmann:2019gfc}.
	
	This paper is organized as follows. In Sec.~\ref{Sec:SMEFT} we introduce the SMEFT framework and effective couplings  relevant to our analysis and describe the computations of SM and BSM contributions to single top-quark production and top-quark decays. In Sec.~\ref{Sec:EFTfitter} we discuss the methodology of our analysis and the experimental input.
	We consider different scenarios for correlations between measurements and demonstrate how such correlations affect the results of fits to current data. 
	Furthermore we study the impact of correlations for future  high-luminosity experiments.
	In Sec.~\ref{Sec:Summary} we conclude. 
	
	\section{SMEFT approach to top-quark physics} 
	\label{Sec:SMEFT}
	
	The SMEFT Lagrangian $\mathcal{L}_\textmd{eff}$ is organized as an expansion in powers of $\Lambda^{-1}$.
	Higher dimensional operators $O_i^{(d)}$ of dimension $d$ are added to the SM Lagrangian $\mathcal{L}_\textmd{SM}$ together with a corresponding Wilson coefficient $C_i^{(d)}$ and a factor $\Lambda^{d-4}$. 
	Since we assume that a gap exist between the scale of new physics and the electroweak scale, we keep only the leading contributions at  $\mathcal{O}(\Lambda^{-2})$:
	\begin{align}
	\mathcal{L}_\textmd{eff}=\mathcal{L}_\textmd{SM}+\sum_i\frac{C^{(6)}_i}{\Lambda^2}O_i^{(6)}+\mathcal{O}\left(\Lambda^{-4}\right)\,.
	\label{Glg:L_eff}
	\end{align}
	In the following we  study $t$-channel single top-quark production cross sections and top-quark decay observables. 
	The following operators contribute at $\mathcal{O}(\Lambda^{-2})$:
	\begin{align}
	\begin{aligned}
	O_{\phi q}^{(3)}&=i\left(\phi^\dagger \overleftrightarrow{D}_{\mu}^{I}\phi\right)\left(\bar{q}_L\gamma^\mu\tau^I q_L\right)\,,\\
	O_{tW}&=\left(\bar{q}_L\sigma^{\mu\nu}\tau^{I}t_R\right)\tilde{\phi}W_{\mu\nu}^{I}	\,,\\
	O_{qq}^{(1)}&=\left(\bar{q}_L\gamma_\mu q_L\right)\left(\bar{q}_L\gamma^\mu q_L\right)	\,,\\
	O_{qq}^{(3)}&=\left(\bar{q}_L\gamma_\mu\tau^I q_L\right)\left(\bar{q}_L\gamma^\mu\tau^I q_L\right)	\,,
	\label{Glg:SingleTopOperator}
	\end{aligned}
	\end{align}
	where $\phi$ denotes the Higgs field, $\tilde \phi_{i} =\epsilon_{ij}\phi_{j}^*$ ($\epsilon_{12}=1$), $t_R$ the top-quark $SU(2)$ singlet and 
	$W^{I}_{\mu\nu}$ and $\tau^I$ are the field strength tensor and the generators of $SU(2)_L$, respectively.
	$q_L$ denotes $SU(2)_L$ doublet quarks of the third generation in $O^{(3)}_{\phi q}$, $O_{tW}$ and first and third generation ones in $O^{(1)}_{qq}$, $O^{(3)}_{qq}$ as specified in the additional superscripts of their Wilson coefficients in Eq.~\eqref{eq:WCs}.
	Neglecting contributions proportional to masses $m\ll m_t$, where $m_t$ is the top-quark mass, the observables depend on three coefficients:
	\begin{align} \label{eq:WCs}
	\tilde C^{(3)}_{\phi q}\,,\  \tilde C_{tW}\,,\  \tilde C_{qq}=\tilde C_{q q}^{(3)1133}+\frac{1}{6}\left(\tilde C_{q q}^{(1)1331}-\tilde C_{q q}^{(3)1331}\right)\,,
	\end{align}
	with $\tilde C_i= C_i v^2/\Lambda^2$ and the Higgs vacuum expectation value $v=\SI{246}{\giga\electronvolt}$. 
	
	In Sec.~\ref{sec:single} and Sec.~\ref{sec:decay} we discuss single top-quark production and top-quark decay, respectively.
	
	\subsection{\boldmath{$t$}-channel single top-quark production \label{sec:single}}
	We employ the Monte Carlo generator \texttt{MadGraph5} \cite{Alwall:2014hca} and the \texttt{dim6top\_LO} UFO model \cite{AguilarSaavedra:2018nen} to compute SM and BSM contributions 
	to total and differential cross sections of $t$-channel single top-quark production at parton level at leading order. 
	We validate our results with \texttt{PYTHIA 8} \cite{Sjostrand:2006za,Sjostrand:2007gs} and find good agreement. 
	For all computations we utilize the MSTW20008lo \cite{Martin:2009iq} parton distribution function (PDF) set.
	To reduce the impact of higher order QCD corrections we take into account SM cross sections at NLO. 
	For differential cross sections we apply $k$-factors to the SM predictions using the NLO predictions presented in the experimental analyses in 
	Refs.~\cite{Aad:2014fwa,Aaboud:2017pdi,CMS:2014ika,CMS:2016xnv}. 
	We validate the results by computing the observables at NLO applying \texttt{MadGraph5} with different PDF sets: MSTW20008nlo \cite{Martin:2009iq}, CT10nlo \cite{Lai:2010vv}, NNPDF23\_nlo \cite{Ball:2013hta}. 
	We find good agreement for all three PDF sets. 
	Total cross sections are computed at NLO using \texttt{MadGraph5} with the same PDF sets.
	Renormalization and factorization scales are set to $\mu_{R,F}=m_t$ with $m_t=\SI{172.8}{\giga\electronvolt}$. 
	Scale uncertainties are evaluated by varying renormalization and factorization scales independently between $m_t/2\leq\mu_{R,F}\leq2m_t$. We take the maximal variation as the 
	uncertainty. We compute PDF uncertainties with \texttt{MadGraph5} using the same PDF sets. We take the central value as the estimate and the total 1 $\sigma$ range, for which we add statistical, 
	PDF uncertainties and scale variation uncertainties in quadrature, as the theory uncertainty.
	
	\subsection{Top-quark decay  \label{sec:decay}}
	The decay width $\Gamma_t$ of the top quark and the $W$ boson helicitiy fractions $F_i$ are measured by extracting $t\bar{t}$ events in the lepton+jets channel with $t\rightarrow Wb$. 
	Hence, we consider only operators modifying the $Wtb$ vertex in Eq.~\eqref{Glg:SingleTopOperator}, $O_{tW}$ and $O^{(3)}_{\phi q}$. 
	Contributions from other dimension-six operators are proportional to the mass of the bottom quark $m_b$, and hence neglected.
	We include BSM contributions at LO and the SM ones at NNLO \cite{Czarnecki:2010gb,Gao:2012ja}. 	
	
	\section{Studying the impact of correlations}
	\label{Sec:EFTfitter}
	We employ a new implementation of the EFT$fitter$ tool \cite{Castro:2016jjv} based on the \textit{Bayesian Analysis Toolkit - BAT.jl} \cite{BAT, BAT.jl} to determine constraints on the Wilson coefficients using a Bayesian ansatz. We include data from 
	ATLAS \cite{Aad:2014fwa,Aaboud:2017pdi,Aaboud:2016ymp,Aaboud:2017uqq,Aad:2012ky,Aaboud:2016hsq}, 
	CMS \cite{Chatrchyan:2012ep,Khachatryan:2014iya,Sirunyan:2016cdg,CMS:2014ika,CMS:2016xnv,Chatrchyan:2013jna,Khachatryan:2014vma,Khachatryan:2016fky}, 
	CDF and D0 \cite{Aaltonen:2010ha,Aaltonen:2012lua,Abazov:2010jn}, given in Tab.~\ref{Tab:Data}.
	\begin{table*}[htbp]
		\centering
		\caption{The experimental measurements of top-quark production and decay considered in this analysis. For both processes we 
			indicate the center of mass energy $\sqrt{s}$, the integrated luminosity, the experiment, the observables included in the analysis and the publication reference.}
		\label{Tab:Data}
		\begin{ruledtabular}
			\begin{tabular}{cccccc}
				Process  & $\sqrt{s}$        &  Luminosity  &  Experiment  & Observable  & Reference\\
				\hline
				Single top  & $\SI{7}{\tera\electronvolt}$  & \makecell{$\SI{4.59}{\femto\barn}^{-1}$\\$\SI{1.17}{\femto\barn}^{-1}$($\mu$)\\ $\SI{1.56}{\femto\barn}^{-1}$(e)}	& 	\makecell{ATLAS\\CMS\\CMS}       & \makecell{$\sigma(tq)$, $\sigma(\bar{t}q)$, ${d\sigma(tq)}/{dp_T}$, ${d\sigma(\bar tq)}/{dp_T}$
					\\$\sigma(tq+\bar{t}q)$\\$\sigma(tq+\bar{t}q)$}
				& \makecell{\cite{Aad:2014fwa}\\\cite{Chatrchyan:2012ep}\\\cite{Chatrchyan:2012ep}}\\
				\hline
				Single top  & $\SI{8}{\tera\electronvolt}$	& \makecell{$\SI{20.2}{\femto\barn}^{-1}$\\$\SI{19.7}{\femto\barn}^{-1}$}	
				& \makecell{ATLAS\\CMS}       & \makecell{$\sigma(tq)$, $\sigma(\bar{t}q)$, ${d\sigma(tq)}/{dp_T}$, ${d\sigma(\bar tq)}/{dp_T}$
					\\$\sigma(tq)$, $\sigma(\bar{t}q)$, $\sigma(tq+\bar{t}q)$, ${d\sigma}/{d|y(t/\bar t)|}$}
				& \makecell{\cite{Aaboud:2017pdi}\\\cite{Khachatryan:2014iya,CMS:2014ika}}\\
				\hline
				Single top  & $\SI{13}{\tera\electronvolt}$	&\makecell{$\SI{3.2}{\femto\barn}^{-1}$\\$\SI{2.2}{\femto\barn}^{-1}$\\ $\SI{2.3}{\femto\barn}^{-1}$}	& \makecell{ATLAS\\CMS\\CMS}       
				& \makecell{$\sigma(tq)$, $\sigma(\bar{t}q)$\\$\sigma(tq)$, $\sigma(\bar{t}q)$, $\sigma(tq+\bar{t}q)$\\ ${d\sigma}/{d|y(t/\bar t)|}$}
				& \makecell{\cite{Aaboud:2016ymp}\\\cite{Sirunyan:2016cdg}\\\cite{CMS:2016xnv}}\\
				\hline
				Top decay &  $\SI{1.96}{\tera\electronvolt}$ &\makecell{$\SI{2.7}{\femto\barn}^{-1}$\\$\SI{8.7}{\femto\barn}^{-1}$\\ $\SI{5.4}{\femto\barn}^{-1}$}	& \makecell{CDF\\CDF\\D0}  
				& \makecell{$F_0$\\$F_0$\\$F_0$}  &  \makecell{\cite{Aaltonen:2010ha}\\\cite{Aaltonen:2012lua}\\\cite{Abazov:2010jn}}\\
				\hline
				Top decay &  $\SI{7}{\tera\electronvolt}$ &\makecell{$\SI{1.04}{\femto\barn}^{-1}$\\$\SI{5.0}{\femto\barn}^{-1}$} &  \makecell{ATLAS\\CMS} 
				& \makecell{$F_0$, $F_L$\\$F_0$, $F_L$} & \makecell{\cite{Aad:2012ky}\\\cite{Chatrchyan:2013jna}}\\
				\hline
				Top decay &  $\SI{8}{\tera\electronvolt}$ &\makecell{$\SI{20.2}{\femto\barn}^{-1}$\\$\SI{20.2}{\femto\barn}^{-1}$\\ $\SI{19.7}{\femto\barn}^{-1}$}&  \makecell{ATLAS\\ATLAS\\CMS} & \makecell{$\Gamma_t$\\ $F_0$, $F_L$\\$F_0$, $F_L$} & \makecell{\cite{Aaboud:2017uqq}\\\cite{Aaboud:2016hsq}\\\cite{Khachatryan:2014vma}}\\
				\hline
				Top decay &  $\SI{13}{\tera\electronvolt}$ & $\SI{19.8}{\femto\barn}^{-1}$& CMS & $F_0$, $F_L$  & \cite{Khachatryan:2016fky}\\
			\end{tabular}
		\end{ruledtabular} 
	\end{table*}
	We count each bin of differential distributions as one observable and include in total 55 measurements of 41 different observables. If differential cross sections are presented 
	in terms of normalized distributions, we reconstruct absolute distributions using total cross sections. We take a constant prior for the parameter interval 
	$-1\leq\tilde{C}_i\leq1$ as default.
	
	We consider both a linear and quadratic fit ansatz for the observables. For the example of total cross sections, the linear one reads
	\begin{align} \label{eq:linear}
	\sigma=\sigma_\textmd{SM}+\sum_i\tilde{C}_i\sigma_i 
	\,, \quad \quad \mbox{('linear')} \, , 
	\end{align}
	where  $\sigma_\textmd{SM}$ denotes the SM contribution and  $\sigma_i$ are the LO interference terms at ${\cal{O}}(1/\Lambda^2)$ between SM and BSM.
	Specifically, in the linear ansatz, the quadratic BSM terms following from squaring amplitudes linear in the Wilson coefficients, are omitted, as they are formally of higher order, $\mathcal{O}(\Lambda^{-4})$, even though they are induced by dimension six operators.
	
	The quadratic ansatz reads
	\begin{align} \label{eq:quadratic}
	\sigma=\sigma_\textmd{SM}+\sum_i\tilde{C}_i\sigma_i + \sum_{i\leq j}\tilde C_i\tilde C_j \sigma_{ij}\,,  \quad \quad \mbox{('quadratic')} \, , 
	\end{align}
	where the purely BSM contributions  from dimension six operators $\sigma_{ij}$ contributing at ${\cal{O}}(1/\Lambda^4)$ are kept.
	To study the performance of the SMEFT-fit in view of the power corrections $v^2/\Lambda^2$, we compare  results in the linear, the quadratic approximation 
	and in a third EFT-implementation
	('linear$+\delta_\text{EFT}$') based on the linear ansatz where we add an additional relative systematic theory uncertainty $\delta_\text{EFT}$
	to each observable to model higher order effects. 
	For each observable, the numerical value of this uncertainty corresponds to its measured value times $\delta_\text{EFT}\sim v^2/\Lambda^2 \approx 0.06$, where we conservatively consider $\Lambda$ as low as $\SI{1}{\tera\electronvolt}$.
	
	In Sec.~\ref{sec:set-up} we provide our set-up for correlated uncertainties. Fit results for present and hypothetical future data are presented in Sec.~\ref{Sec:Compare} and Sec.~\ref{Sec:Future}, respectively.
	In Sec.~\ref{sec:lit} we compare our findings assuming no correlations to results in  the literature.
	
	\subsection{Uncertainty set-ups \label{sec:set-up}}
	We consider three different types of uncertainties: Statistical uncertainties, systematic uncertainties and theory uncertainties.
	In the statistical analysis with EFT\emph{fitter} the uncertainties of all measurements are assumed to be Gaussian distributed. As described in 
	Ref.~\cite{Castro:2016jjv}, correlations are taken into account for all types of uncertainties (here: statistical, systematic and theory) by calculating the total covariance matrix $\mathcal{M}$ as the sum of the 
	individual covariance matrices $\text{cov}^{(k)}[x_i, x_j]$
	\begin{align}
	\mathcal{M}_{ij} = \text{cov}[x_i, x_j] = \sum_k \text{cov}^{(k)}[x_i, x_j] = \sum_k \rho^{(k)}_{ij} \sigma_i^{(k)} \sigma_j^{(k)}\,,
	\end{align}
	where $x_i$ denotes the measurements, $\sigma_i^{(k)}$ are the uncertainty values and $\rho_{ij}^{(k)}$ are the correlation coefficients and the sum is over all types of uncertainties.
	Correlated statistical uncertainties arise if different observables are extracted from the same data set.
	Corresponding correlation matrices are mostly known from the experimental analyses \cite{Aad:2014fwa, Aaboud:2017pdi, Aad:2012ky,Aaboud:2016hsq,Khachatryan:2014vma,Khachatryan:2016fky}, and included in our analysis.
	In contrast, almost no information about the correlation of systematic uncertainties or theory uncertainties is provided. 
	To study their impact on the results of the fit 
	we choose a simple parametrization of the correlation matrices. 
	In particular, we model the overall impact using effective correlation coefficients covering all sources of systematic and theory uncertainties, respectively.
	As described in more detail below, the choice of the overall correlation coefficients is motivated by the combination of ATLAS and CMS single-top quark production cross-sections in Ref.~\cite{Aaboud:2019pkc}.
	
	In the case of systematic uncertainties, 
	correlations between measurements by the same experiment at the same energy are set to $\rho_\text{sys}$, since such systematic uncertainties are expected to have the same source.
		Moreover, we expect the uncertainties of observables measured by the same experiment at different energies to be correlated less, and therefore set these entries to $\rho_\text{sys}/2$. 
	In particular,  all bins of differential distributions are considered with the same correlation coefficient. 
		Since almost all of the distributions included in the fit are unnormalized, correlations beyond neighbouring bins can appear and the situation becomes very complicated. 
		Taking  these effects into account realistically requires input from the experiments, and should ideally be provided.
	
	In contrast, theory uncertainties are not expected to depend on the experiment, but  on the energy of the process. 
	Therefore, correlations between measurements at the same energy are set to $\rho_\text{th}$. Observables measured at different energies are assumed to be correlated with a coefficient 
	$\rho_\text{th}/2$. 
	In general, the size of theory correlations depends on the energy of the collision. However, without further inputs from the experimental collaborations, it is not possible to provide proper estimates of the correlation coefficients. In our analysis, we find that the effect of these correlations is small relative to the systematic ones. Thus, we choose a simple approach with only two kinds of correlation 
	coefficients, $\rho_\text{th}$ and $\rho_\text{th}/2$.
	 In the fit with the linear$+\delta_\text{EFT}$ ansatz we include an additional correlation matrix to model an EFT uncertainty. For this matrix we make the same assumptions as for the theory uncertainties, {\it i.e.}, take  $\rho_\text{EFT}$ and $\rho_\text{EFT}/2$, with $\rho_\text{EFT}=0.9$.
	
	It should be noted that $F_0$ and $F_L$ are always anti-correlated in our set-up since they are required to add up to $1-F_R$.
	In the SM, $F_R=0+\mathcal{O}(m_b^2/m_t^2)$ due to the $V-A$ structure of the weak interaction. Contributions 
	from $O_{tW}$ are suppressed by a factor $m_b^2/m_t^2$ and contributions from additional dimension-six operators including right-handed bottom quarks are suppressed 
	by a factor $m_b/m_t$. We neglect these contributions in the fit.
	
	In the following analysis we demonstrate the impact of the correlation parameters on the fit results by varying $\rho_\text{sys}$ and $\rho_\text{th}$ independently within
	the interval $[0,1]$, 
	since positive values for the correlations are expected. We also explored the possibility of negative values  but found that in this case the covariance matrix is no longer positive definite for $\rho_\text{sys,th} < -0.075$.
	We present results for two benchmark scenarios: The 'no correlation' scenario, which has been adopted in previous studies \cite{Buckley:2015lku,Brown:2019pzx,Hartland:2019bjb}, where we neglect
	all unknown correlations 
	\begin{align}  \label{eq:nc}
	\rho_\text{sys}=\rho_\text{th}=0 \, , \quad \quad (\mbox{'no correlation'})
	\end{align}
	and the  'best guess' scenario with strong correlations \cite{Aaboud:2019pkc} 
	\begin{align}
	\rho_\text{sys}=0.9\,,\quad \rho_\text{th}=0.9\,, \quad \quad (\mbox{'best guess'}) \, . 
	\label{Eq:bestguess}
	\end{align}
	Our 'best guess' scenario is supported by a similar combination of measurements in Ref.~\cite{Aaboud:2019pkc}.
	In this combination, all correlations of systematic uncertainties, except for the integrated luminosity, between measurements of ATLAS and CMS are set to zero,
	while theory uncertainties are assumed to be maximally correlated.
	For these reasons, we assume in our analysis systematic uncertainties between CMS and ATLAS to be uncorrelated.
	We have checked explicitly  that their impact is negligible given present uncertainties.
	Specifically, we investigated the impact of replacing all zeros in the ‘best guess’ correlation matrix of systematic uncertainties by $\rho_z$, which then is varied between -0.25 and 0.25. Corresponding shifts in the Wilson coefficients lie within the $95\,\%$ intervals and are negligible.
	
	For theory uncertainties of observables at the same energy we assume strong correlation coefficients of $0.9$. We do not assume maximal correlations, since the effect of theory uncertainties depends also on details of the experimental analyses. 
	For measurements by the same experiment at different energies,
	theory uncertainties are assumed to be less strongly correlated due to the energy dependency of the theory modelling.  
	Observables measured in different processes or channels are independent and hence are assumed to have uncorrelated theory and systematic uncertainties for simplicity. 
	It should be noted that we do not address the problem that background processes are also affected by EFT operators, as this would require a reanalysis of the measurements that have been performed using SM assumptions.
	
	The correlation matrices for the data given in Tab.~\ref{Tab:Data} are $55\times55$ dimensional and too large to be given here explicitly. 
	Instead, we illustrate our parametrization with a simplified one. Suppose a dataset with five measurements:
	the total cross sections of single top-quark production $\sigma(tq)^\textmd{A}_{7}$ and single antitop-quark production $\sigma(\bar tq)^\textmd{A}_{7}$ performed by ATLAS at $\SI{7}{\tera\electronvolt}$, 
	the total cross section $ \sigma(tq)^\textmd{A}_{8}$ and $\sigma(tq)^\textmd{C}_{8}$ measured at $\SI{8}{\tera\electronvolt}$ by ATLAS and CMS, respectively, and 
	the top-quark decay width $\Gamma_t$.
	In this example, our parametrization of the correlation matrix of systematic uncertainties reads
	\begin{align}
	\bordermatrix{
		&  \sigma(tq)^\textmd{A}_{7}   	&  \sigma(\bar{t}q)^\textmd{A}_{7} 	& \sigma(tq)^\textmd{A}_{8} 	&	 \sigma(tq)^\textmd{C}_{8} 	&	\Gamma_t\cr 
		\sigma(tq)^\textmd{A}_{7} 	& 	1 			&  {\rho_\text{sys}} 		&  {\frac{\rho_\text{sys}}{2}}	 	&		0			&		0\cr
		\sigma(\bar{t}q)^\textmd{A}_{7}	& 	{\rho_\text{sys}} 	& 	1 				&  {\frac{\rho_\text{sys}}{2}} 	&		0			&		0	\cr
		\sigma(tq)^\textmd{A}_{8} 	&  	{\frac{\rho_\text{sys}}{2}} 	&  	{\frac{\rho_\text{sys}}{2}} 		& 	1			&		0			&		0\cr
		\sigma(tq)^\textmd{C}_{8}	&	0			&		0			&	0			&		1			&		0\cr
		\Gamma_t			&	0			&		0			&	0			&		0			&		1\cr
	}\,,
	\label{Glg:CorrMatrix_sys}
	\end{align}
	while the one of theory uncertainties is written as
	\begin{align}
	\bordermatrix{
		&  \sigma(tq)^\textmd{A}_{7}   	&  \sigma(\bar{t}q)^\textmd{A}_{7} 	& \sigma(tq)^\textmd{A}_{8} 	&	 \sigma(tq)^\textmd{C}_{8} 	&	\Gamma_t\cr 
		\sigma(tq)^\textmd{A}_{7} 	& 	1 			&  {\rho_\text{th}} 		&  {\frac{\rho_\text{th}}{2}}	 	&		{\frac{\rho_\text{th}}{2}}			&		0\cr
		\sigma(\bar{t}q)^\textmd{A}_{7}	& 	{\rho_\text{th}} 	& 	1 				&  {\frac{\rho_\text{th}}{2}} 	&		{\frac{\rho_\text{th}}{2}}		&		0	\cr
		\sigma(tq)^\textmd{A}_{8} 	&  	{\frac{\rho_\text{th}}{2}} 	&  	{\frac{\rho_\text{th}}{2}} 		& 	1			&		\rho_\text{th}			&		0\cr
		\sigma(tq)^\textmd{C}_{8}	&	{\frac{\rho_\text{th}}{2}}			&		{\frac{\rho_\text{th}}{2}}			&	\rho_\text{th}			&		1			&		0\cr
		\Gamma_t			&	0			&		0			&	0			&		0			&		1\cr
	}\,.
	\label{Glg:CorrMatrix_theo}
	\end{align}
	The additional matrix of the $\delta_\text{EFT}$ uncertainties included in the linear$+\delta_\text{EFT}$ 
	ansatz reads
	\begin{align}
	\bordermatrix{
		&  \sigma(tq)^\textmd{A}_{7}   	&  \sigma(\bar{t}q)^\textmd{A}_{7} 	& \sigma(tq)^\textmd{A}_{8} 	&	 \sigma(tq)^\textmd{C}_{8} 	&	\Gamma_t\cr 
		\sigma(tq)^\textmd{A}_{7} 	& 	1 			&  \rho_\text{EFT} 		&  \frac{\rho_\text{EFT}}{2}	 	&		\frac{\rho_\text{EFT}}{2}			&		0\cr
		\sigma(\bar{t}q)^\textmd{A}_{7}	& 	\rho_\text{EFT} 	& 	1 				&  \frac{\rho_\text{EFT}}{2} 	&		\frac{\rho_\text{EFT}}{2}		&		0	\cr
		\sigma(tq)^\textmd{A}_{8} 	&  	\frac{\rho_\text{EFT}}{2} 	&  	\frac{\rho_\text{EFT}}{2} 		& 	1			&		\rho_\text{EFT}			&		0\cr
		\sigma(tq)^\textmd{C}_{8}	&	\frac{\rho_\text{EFT}}{2}			&		\frac{\rho_\text{EFT}}{2}		&	\rho_\text{EFT}			&		1			&		0\cr
		\Gamma_t			&	0			&		0			&	0			&		0			&		1\cr
	}\,,
	\label{Glg:CorrMatrix_EFT}
	\end{align}
	with $\rho_\text{EFT}=0.9$ in both the 'no correlation' and 'best guess' scenario. 
	We do not include correlations between observables from different processes (in this case single-top production and top decay).
	While single-top processes are affected by all three coefficients ($\tilde C^{(3)}_{\varphi q}$, $\tilde{C}_{tW}$, $\tilde C_{qq}$), 
	contributions $\mathcal{O}(\Lambda^{-2})$ from $\tilde C_{qq}$ to top-quark decay observables are negligible as the $W$ bosons in $t\rightarrow Wb$ are produced on-shell \cite{Brivio:2019ius}. 
	In addition, in measurements of the $W$ boson helicity fractions the $W$ boson is reconstructed from its decay particles. This further suppresses any contributions from four-fermion operators.
	Thus, only the two coefficients $\tilde C^{(3)}_{\varphi q}$ and $\tilde{C}_{tW}$ contribute to top-quark decay observables.
	With the parametrization in Eq.~\eqref{Glg:CorrMatrix_EFT} we aim for a description of the energy dependency of BSM contributions ($\rho_\text{EFT}$, $\rho_\text{EFT}/2$) while a correlation of observables from different processes affected by a different number of operators lies beyond the scope of this ansatz. 


	
	\subsection{Fit to data}
	\label{Sec:Compare}
	\begin{figure}[htbp]
		\begin{center}
			\includegraphics[width=0.49\textwidth]{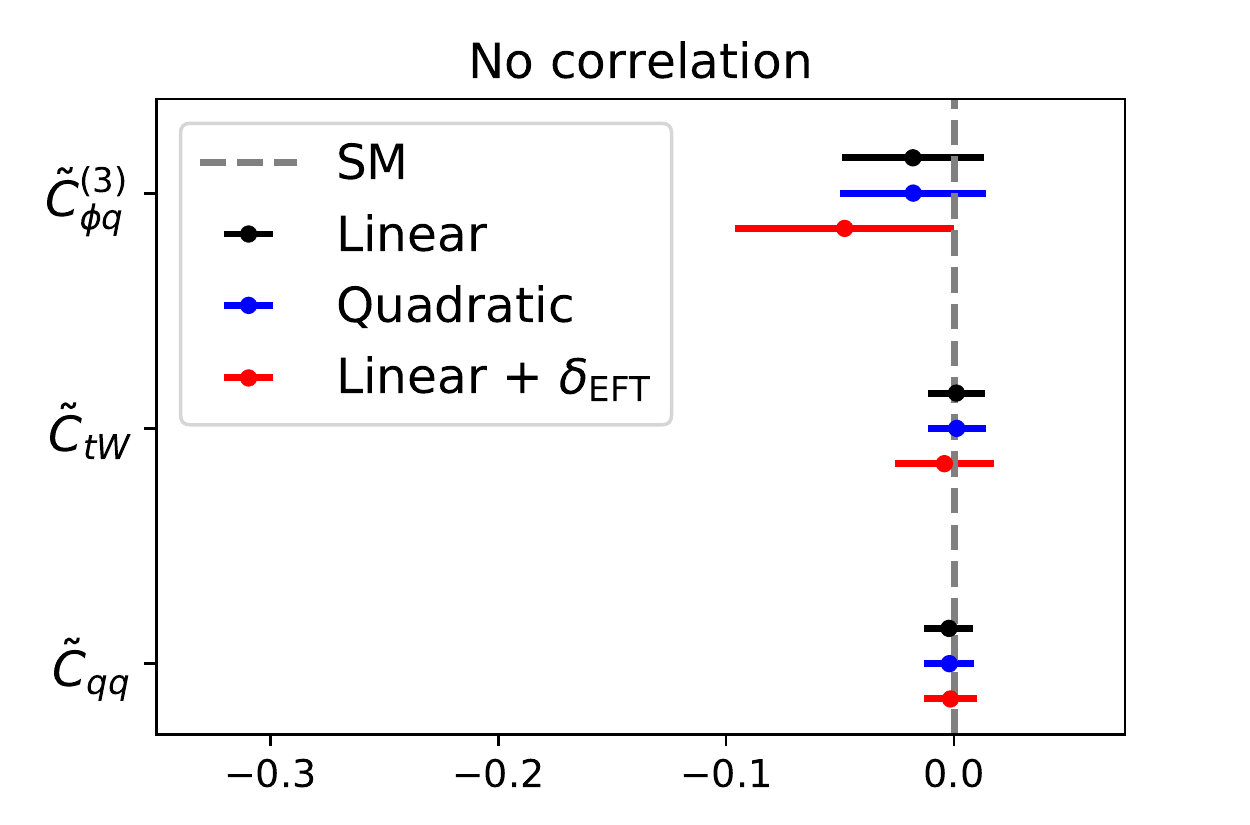}
			\includegraphics[width=0.49\textwidth]{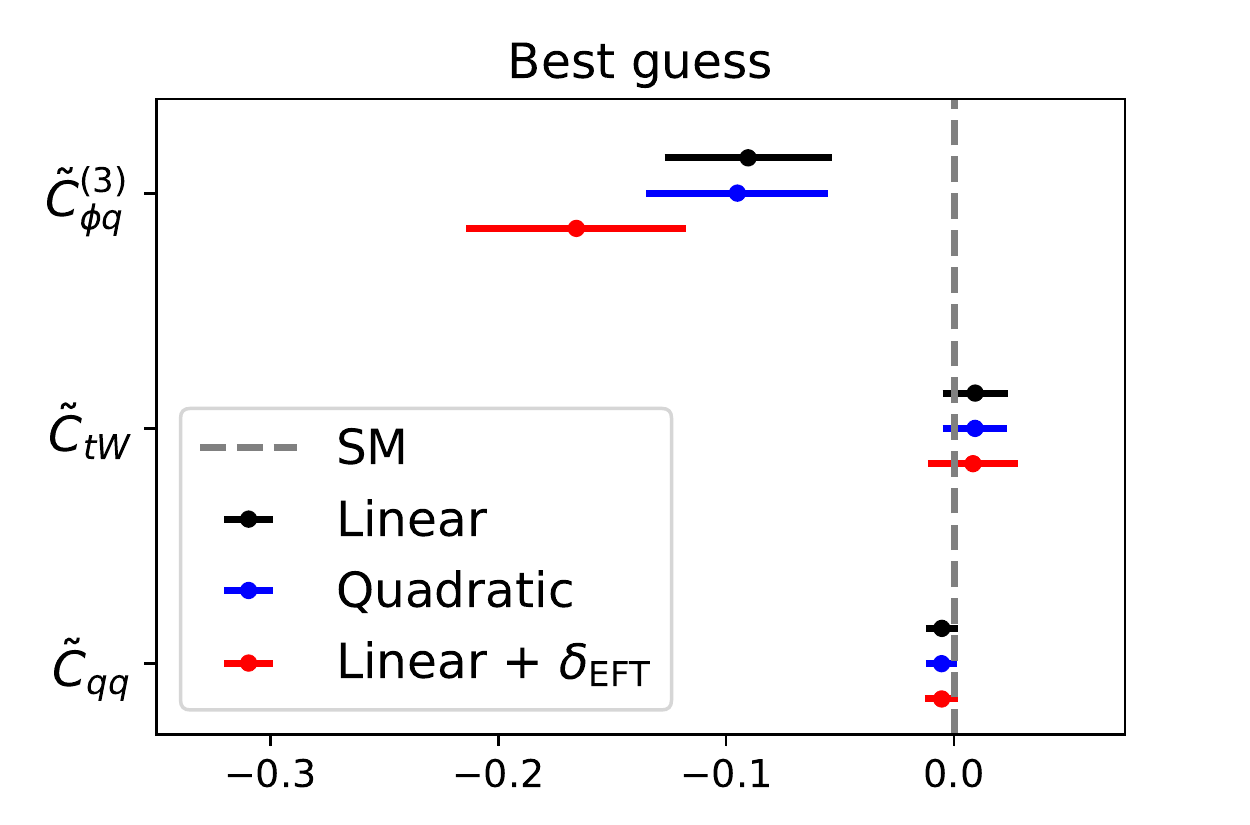}
		\end{center}
		\caption{Marginal constraints on the coefficients $\tilde{C}_i$ from fits with the 'no correlation' scenario, Eq.~\eqref{eq:nc}, and the 'best guess' scenario, Eq.~\eqref{Eq:bestguess}. 
			Dots and lines denote the central value and the smallest $95\,\%$ interval, respectively, in the 1D projection. The SM is indicated by the  vertical dashed line.}
		\label{Fig:Constraints1D}
	\end{figure}
	
	\begin{table*}[ht]
		\centering
		\caption{Marginalized smallest $\SI{95}{\percent}$ intervals obtained in fits in the 'no correlation' scenario Eq.~\eqref{eq:nc} and the 'best guess' scenario Eq.~\eqref{Eq:bestguess} to the data in Tab.~\ref{Tab:Data} shown in Fig.~\ref{Fig:Constraints1D}. The central value is in the center of these intervals.}		
		\label{Tab:Now1D}
		\begin{ruledtabular}
			\begin{tabular}{cccc}
				Operators& Linear  & Linear$+\delta_\text{EFT}$  & Quadratic \\\hline
				'No correlation'	&&&\\\hline
				$\tilde C^{(3)}_{\phi q}$  & $[-0.049,\,0.014]$ & $[-0.096,\,0.000]$  &  $[-0.050,\,0.013]$  \\
				$\tilde C_{tW}$   & $[-0.012,\,0.014]$ & $[-0.026,\,0.018]$ & $[-0.012,\,0.014]$\\
				$\tilde C_{q q}$  & $[-0.013,\,0.009]$ & $[-0.013,\,0.010]$  & $[-0.013,\,0.009]$\\\hline
				'Best guess'	&&&\\\hline
				$\tilde C^{(3)}_{\phi q}$  & $[-0.127,\,-0.055]$ & $[-0.214,\,-0.118]$  &  $[-0.135,\,-0.056]$  \\
				$\tilde C_{tW}$   & $[-0.011,\,0.029]$ & $[-0.012,\,0.029]$ & $[-0.005,\,0.024]$\\
				$\tilde C_{q q}$  & $[-0.012,\,0.002]$ & $[-0.013,\,0.002]$  & $[-0.012,\,0.002]$\\
			\end{tabular}
		\end{ruledtabular}
	\end{table*}

	We present results from fits to the data given in Tab.~\ref{Tab:Data}. The fit is performed with a linear ansatz (\ref{eq:linear}) with and without the additional uncertainty $\delta_\text{EFT}$, and with a quadratic 
	one (\ref{eq:quadratic}). In the following we denote with 'central value' the global mode of the posterior distribution.

	In Fig.~\ref{Fig:Constraints1D} and Tab.~\ref{Tab:Now1D} we give marginal constraints on each coefficient in the 'no correlation' scenario and the 'best guess' scenario.
	For each coefficient we give the central value (dot) and the smallest $95\,\%$ intervals (lines) in the 1D projections. 
	Concerning the different EFT-implementations we find for all three coefficients good agreement between  the fits in the linear and the quadratic ansatz.
	For $\tilde{C}_{tW}$ and $\tilde{C}_{q q}$ we also find that the 'linear + $\delta_\text{EFT}$' shows good agreement with the other two fits. In case of $\tilde C_{tW}$ we see that in the 'linear + $\delta_\text{EFT}$' scenario the 95\,\% interval is in general larger.
	For $\tilde C^{(3)}_{\phi q}$ we find differences in the 'linear + $\delta_\text{EFT}$' compared to the other two scenarios: The 95\,\% interval is larger in both correlation scenarios and also shifted towards negative values away from the SM. In the 'no correlation' scenario the three fits are still in agreement with each other (the deviations compared to linear and quadratic fit are $\sim 1.2~\sigma$) while in the 'best guess' scenario the 'linear + $\delta_\text{EFT}$' deviates by roughly 3~$\sigma$.
	This indicates that the neglected quadratic terms in the linear fit are a negligible source of uncertainty which is in turn overestimated in the 'linear + $\delta_\text{EFT}$' ansatz. 

	Correlations have a significant impact on the results of the fit. While the results for $\tilde{C}_{tW}$ and $\tilde C_{qq}$ are less affected, we find significant changes in the 'best guess' scenario compared to the 'no correlation' scenario for $\tilde{C}^{(3)}_{\phi q}$.
	The central value of $\tilde{C}^{(3)}_{\phi q}$ is shifted away from the SM and the $95\,\%$ interval grows by roughly 10\,\%. This gives rise to deviations from the SM
	of up to $4.7$ $\sigma$ in the linear/quadratic and $6.6$ $\sigma$ in the 'linear + $\delta_\text{EFT}$' ansatz. 
	In the case of $\tilde C_{tW}$ the 95~\% intervals are shifted to positive values. 
	This is due to differential cross section measurements which in the presence of strong correlations lead to positive values of $\tilde C_{tW}$. Similarly, the shrinking of the interval of $\tilde C_{qq}$ is also due to shifts from differential cross sections. These observables dominate constraints on $\tilde C_{qq}$. For strong correlations the 95~\% region derived from differential cross section measurements is deformed in the three-dimensional parameter space resulting in a smaller interval for $\tilde C_{qq}$, as expected for additional correlations.

	\begin{figure*}[]
		\begin{center}
			\includegraphics[width=0.46\textwidth]{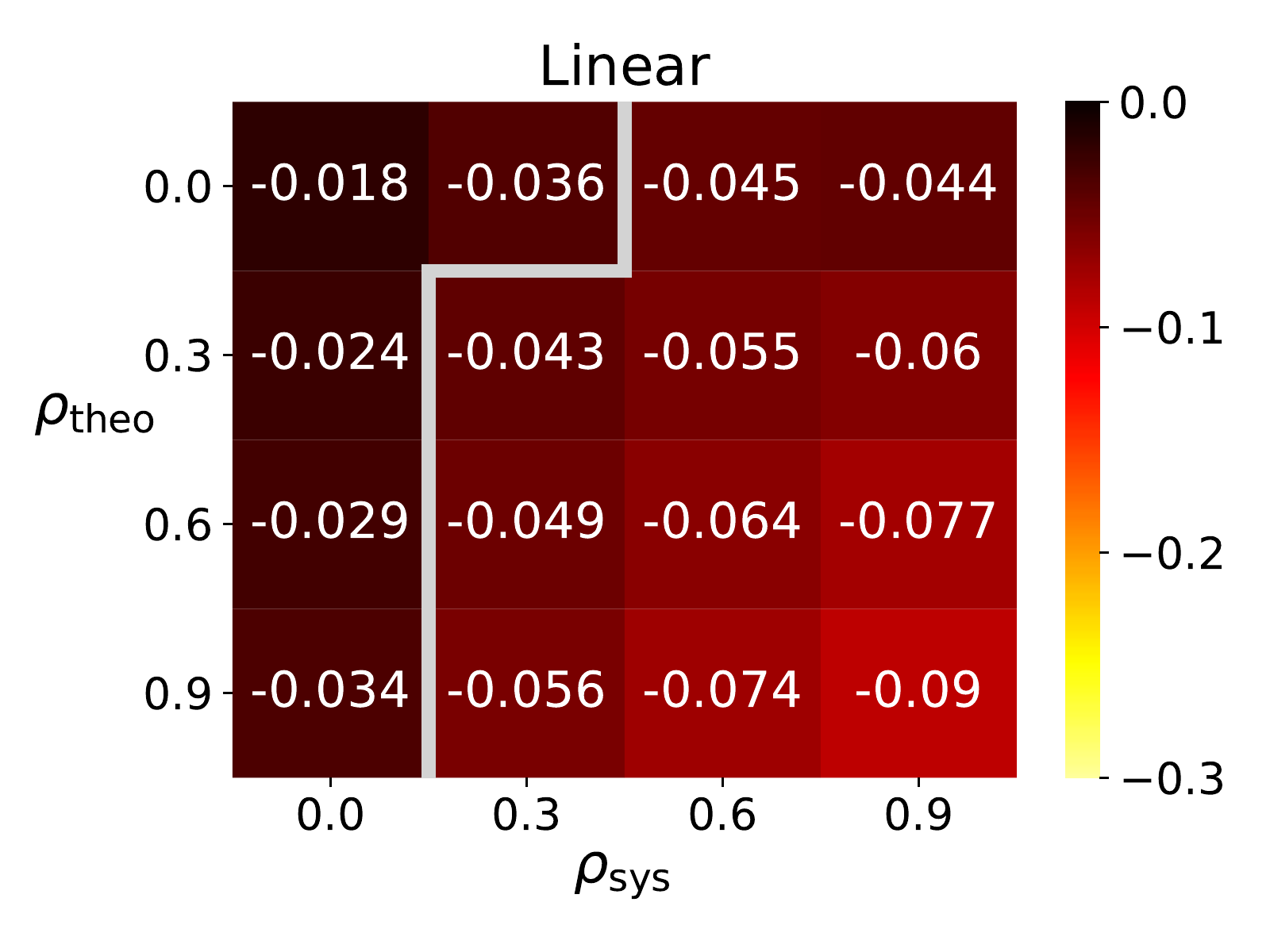}
			\includegraphics[width=0.46\textwidth]{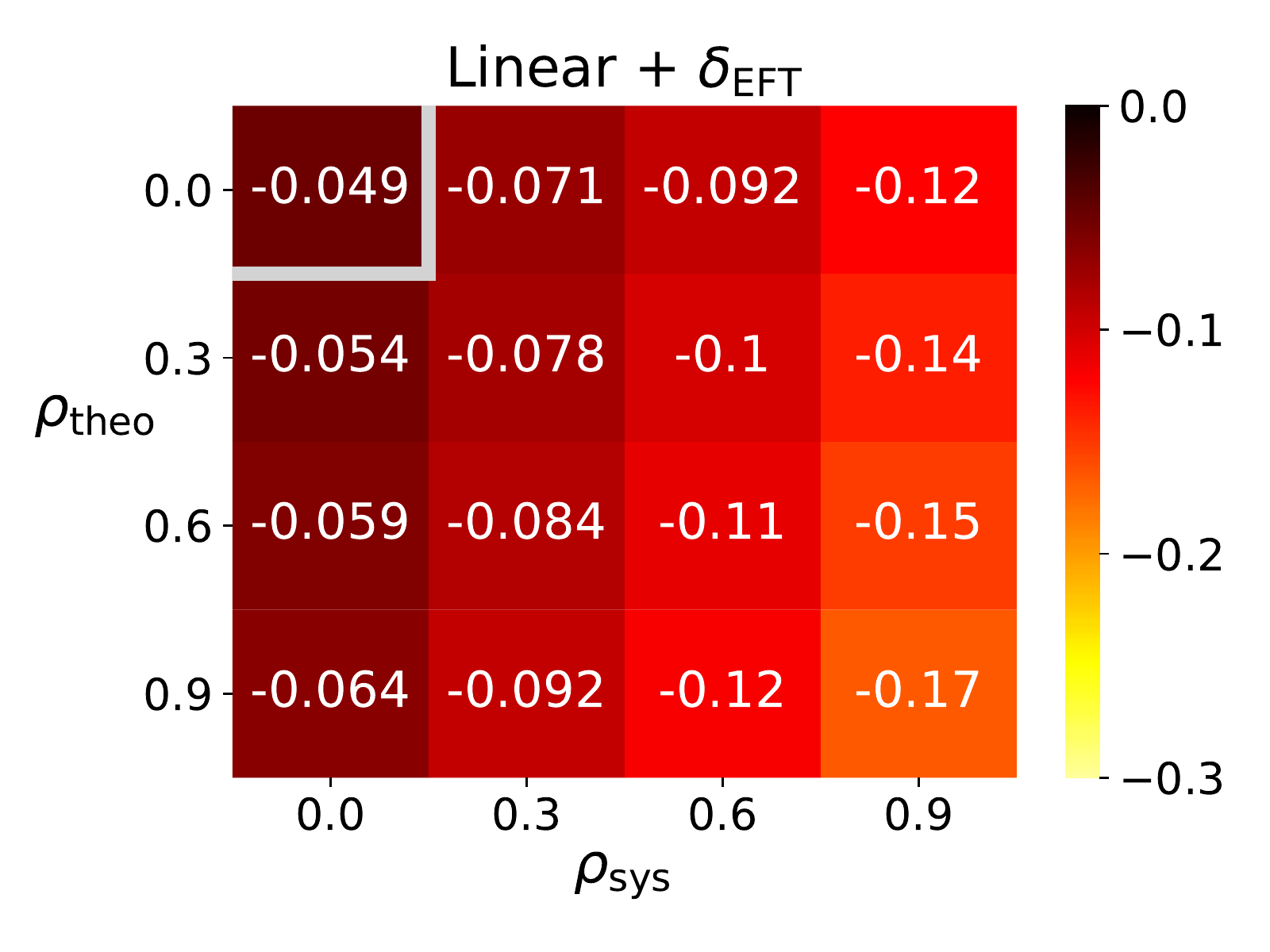}
			\includegraphics[width=0.46\textwidth]{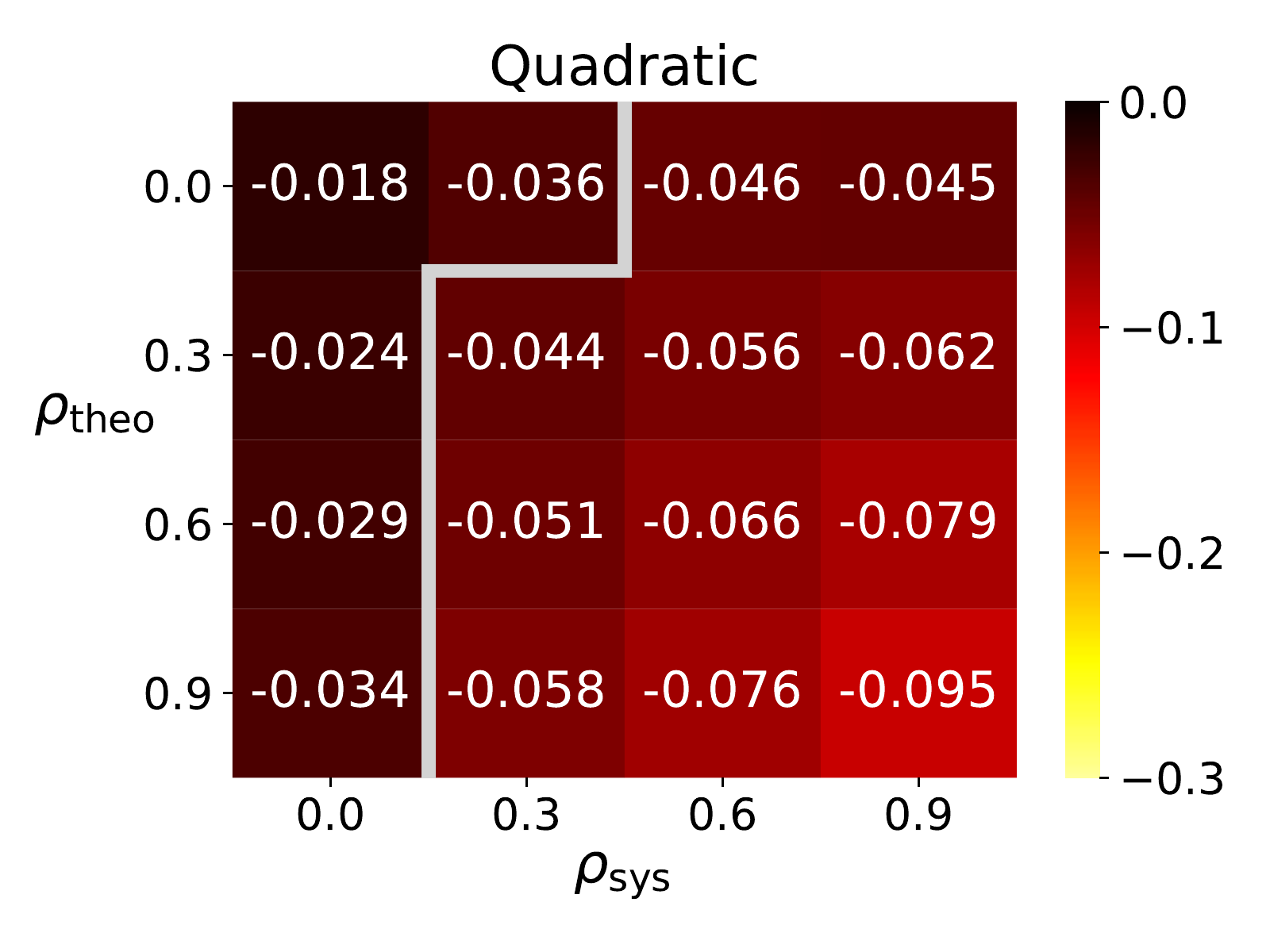}
		\end{center}
		\caption{Central values of $\tilde{C}^{(3)}_{\phi q}$ in the different EFT-implementations for correlation parameters  $\rho_\text{sys},\,\rho_\text{th} = 0.0,\,0.3,\,0.6,\,0.9$ from a
		marginalized fit to the data given in Tab.~\ref{Tab:Data}.
		Both correlation parameters are varied independently from each other. 
		The upper-left and lower-right corner correspond to the 'no correlation' scenario in Eq.~\eqref{eq:nc} and the 'best guess' scenario in Eq.~\eqref{Eq:bestguess}, respectively.
		Central values below and to the right of the grey line are in conflict with the SM at more than $2$ $\sigma$.}
		\label{Fig:Correlation2D_current}
		\begin{center}
			\includegraphics[width=\textwidth]{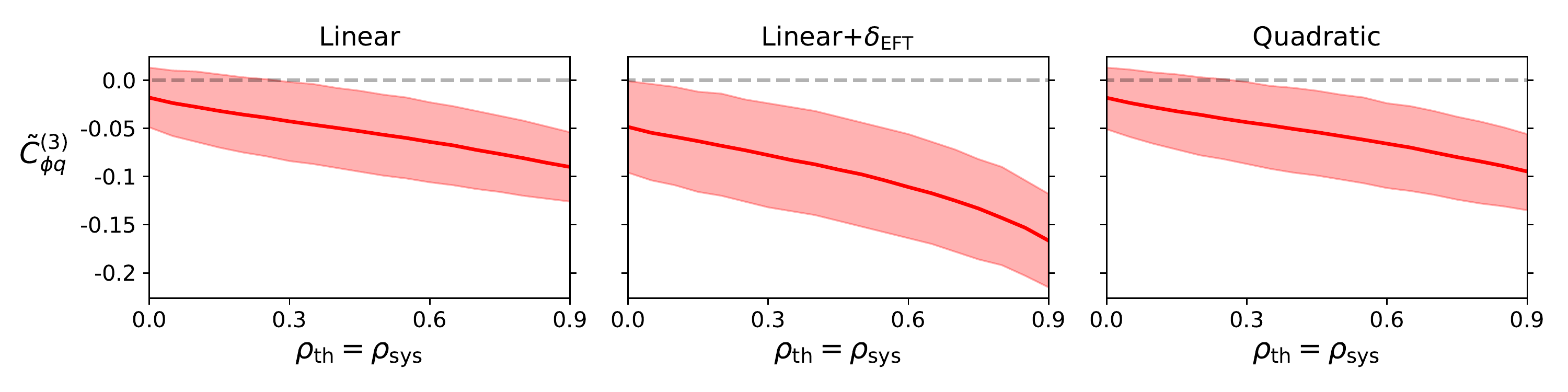}
			\caption{Central values (red line) and $95\,\%$ intervals (red band) of $\tilde{C}^{(3)}_{\phi q}$ in the different EFT-implementations for correlation parameters  
				$\rho_\text{sys}=\rho_\text{th} \in [0,0.9] $ from a marginalized fit to the data given in Tab.~\ref{Tab:Data}.}
			\label{Fig:Scan}
		\end{center}
	\end{figure*}
	To detail the impact of systematic and theory uncertainties on the constraints we perform fits in which we vary $\rho_{\rm sys}$  and $\rho_{\rm th}$  independently.
	In Fig.~\ref{Fig:Correlation2D_current} we give the central value of $\tilde{C}^{(3)}_{\phi q}$ for  $\rho_\text{sys},\,\rho_\text{th} = 0.0,0.3,0.6,0.9$ in the different EFT-implementations.
	The upper-left corner corresponds to the 'no correlation' scenario and the lower-right corner to the 'best guess' scenario. 
	
	The $95\,\%$ region changes with varying correlation coefficients by a factor of up to 1.5 in all three implementations and is not shown for simplicity.
	In linear and quadratic implementation of the BSM contributions we find very similar results while in the 'linear + $\delta_\text{EFT}$' implementation the deviations from the SM are more pronounced, similar to Fig.~\ref{Fig:Constraints1D}.
	Correlations of theory uncertainties and systematic uncertainties affect the constraints in a similar way. With increasing values of  $\rho_i$ the central value is shifted further away from both the SM and the 'no correlation' scenario. 
	The grey line shows which central values deviate strongly from the 'no correlation' scenario: values below and to the right of the line deviate from the SM by more than $2$ $\sigma$.
	We find that correlations of systematic uncertainties have a stronger impact on the constraints than correlations of theory uncertainties since even in the case $\rho_\text{th}=0$ we can still 
	find deviations of more than $2$ $\sigma$ from the SM while we do not find such deviations for $\rho_\text{sys}=0$ in the linear and quadratic models. In the 'linear + $\delta_\text{EFT}$' scenario we observe a similar behavior where the central value deviates stronger from the SM for rising values of $\rho_\text{sys}$ than for $\rho_\text{th}$.

	Correlations affect constraints on $\tilde{C}^{(3)}_{\phi q}$, while $\tilde C_{qq}$ and $\tilde{C}_{tW}$ remain almost unchanged. 
	This is due to the different datasets driving the constraints: $\tilde{C}_{tW}$ is strongly constrained by the helicity fractions, which have weaker correlations among each other and smaller uncertainties than 
	single top production data. In contrast, $\tilde C_{qq}$ and $\tilde{C}^{(3)}_{\phi q}$ are constrained by the differential and total cross sections. These datasets can be strongly correlated, 
	such that the corresponding constraints on the Wilson coefficients can change significantly with the correlation set-up. In the case of $\tilde C_{qq}$ the BSM contributions are significantly larger and the constraints are stronger compared to $\tilde C^{(3)}_{\phi q}$ so that effects of correlations are less pronounced.
	
	In Fig.~\ref{Fig:Scan} we give central values (red line) and the corresponding $95\,\%$ intervals (red band) of $\tilde{C}^{(3)}_{\phi q}$ in the different EFT-implementations for correlation parameters  
	$\rho_\text{sys}=\rho_\text{th} \in [0,0.9] $ from a marginalized fit to the data given in Tab.~\ref{Tab:Data}. We find a continuous and consistent behavior of the constraints for 
	increasing correlation parameters.
	As evident from all Figs.~\ref{Fig:Constraints1D}-\ref{Fig:Scan}, stronger correlations result in stronger deviations from the SM. 
	\begin{figure}[]
		\begin{center}
			\includegraphics[width=0.49\textwidth]{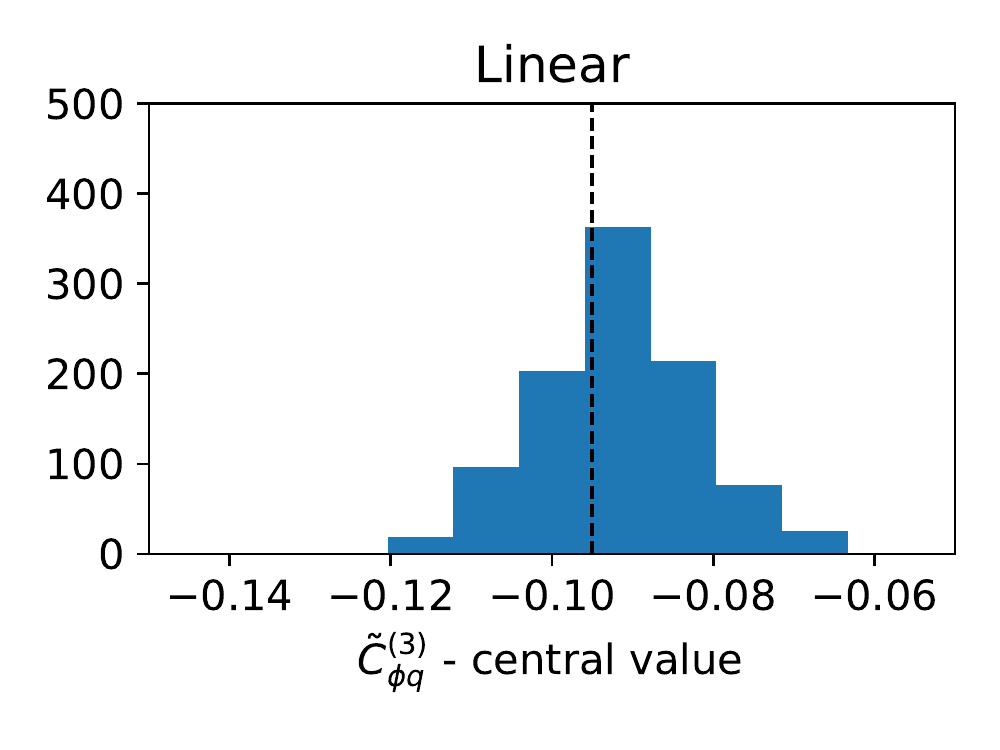}
			\includegraphics[width=0.49\textwidth]{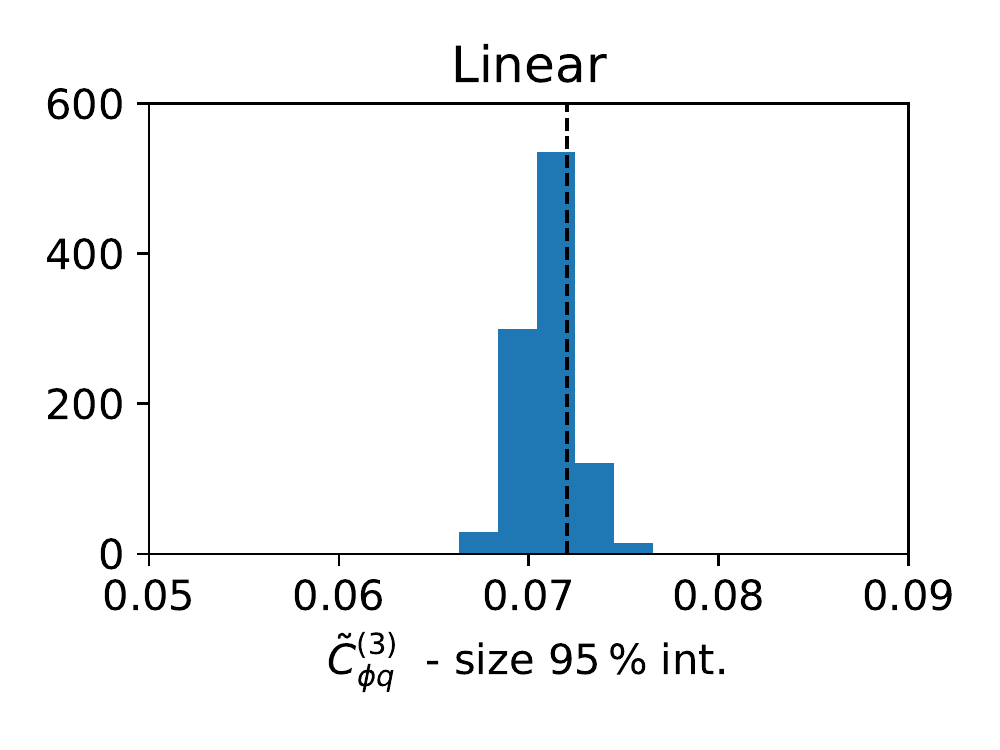}
			\caption{Histograms of the central value and the size of the $95\,\%$ interval of $\tilde{C}^{(3)}_{\phi q}$ in the linear EFT-implementations 
				for correlation parameters varied randomly around the 'best guess' scenario from a marginalized fit to data given in Tab.~\ref{Tab:Data}. Black lines denote results from the 'best guess' scenario 
				in Eq.~\eqref{Eq:bestguess}.}
			\label{Fig:Hist}
		\end{center}
	\end{figure}
	
	To validate the stability of our fit, we vary the non-zero off-diagonal entries in the 'best guess'
	correlation matrices of systematic and theory uncertainties by adding uniformly distributed random numbers $u$ with $|u|\leq0.03$ to the entries. 
	Each element is varied individually while keeping the correlation matrices positive semi-definite. 
	Using the randomized correlation matrices, we perform 1000 marginalized fits to the data given in Tab.~\ref{Tab:Data}.
	In Fig.~\ref{Fig:Hist} we give histograms for the central value and for the size of the $95\,\%$ interval of $\tilde{C}^{(3)}_{\phi q}$ in the linear EFT-implementations 
	for correlation parameters varied randomly around the 'best guess' scenario  (\ref{Eq:bestguess}). The black lines denote the results from the 'best guess' scenario.
	Compared to the 'best guess' scenario, the distribution of the central values is  symmetric and slightly shifted to values closer to the SM.
	The distribution of the size of the $95\,\%$ interval is slightly shifted to smaller values and shows an asymmetry, favoring smaller intervals. 
	As in  Fig.~\ref{Fig:Scan}, we observe a smooth and stable dependence on the correlation parameters.
	Similar results are obtained for the quadratic and linear$+\delta_\text{EFT}$ EFT-implementations (not shown).
	
	\subsection{Future scenarios} \label{Sec:Future}
	We demonstrate the impact of correlations of systematic and theory uncertainties in the light of the higher integrated luminosity at future experiments, such as LHC Run-3 and HL-LHC \cite{ApollinariG.:2017ojx}, 
	considering different future scenarios:
	$300\,\textmd{fb}^{-1}$ and $3000\,\textmd{fb}^{-1}$. To do so, we scale statistical uncertainties of the data in Tab.~\ref{Tab:Data} 
	according to the presumed integrated luminosity keeping the present central values and systematic and theory uncertainties.
	\begin{figure}[h]
		\begin{center}
			\includegraphics[width=0.49\textwidth]{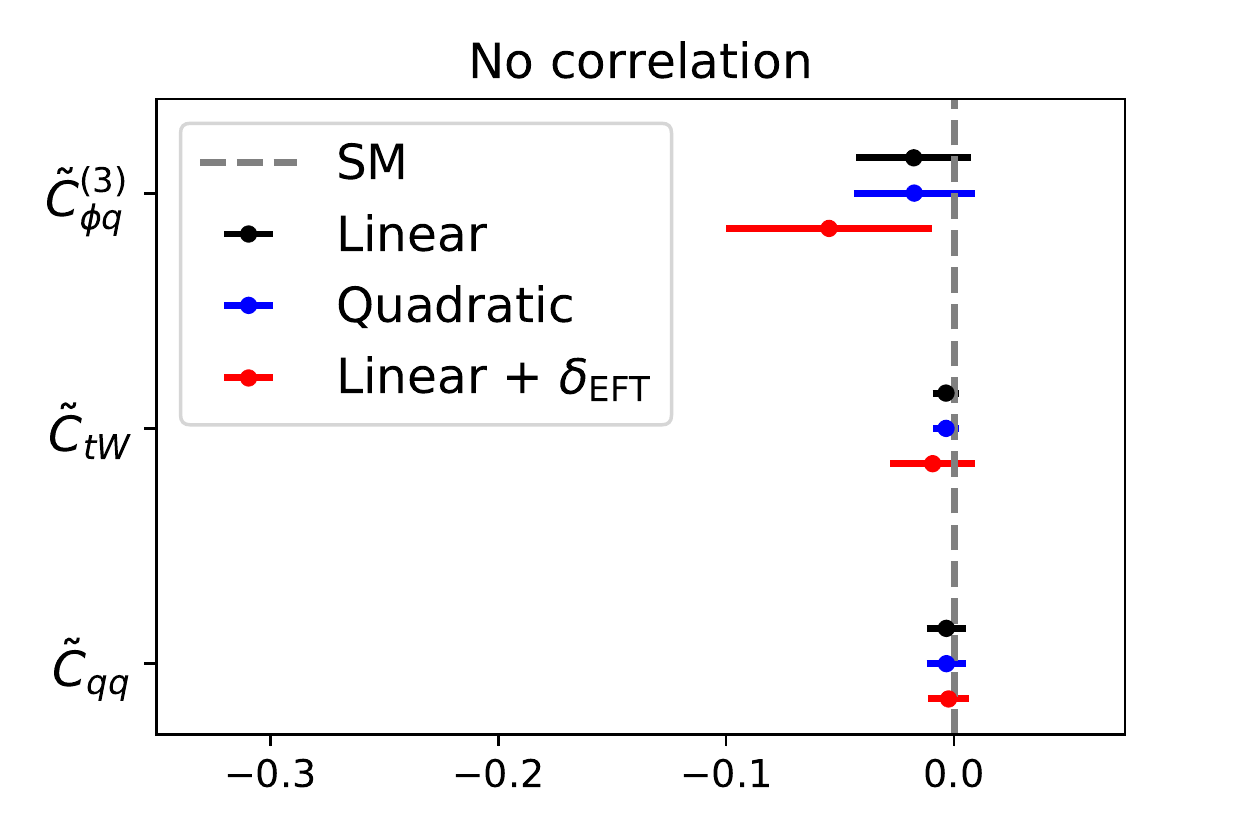}
			\includegraphics[width=0.49\textwidth]{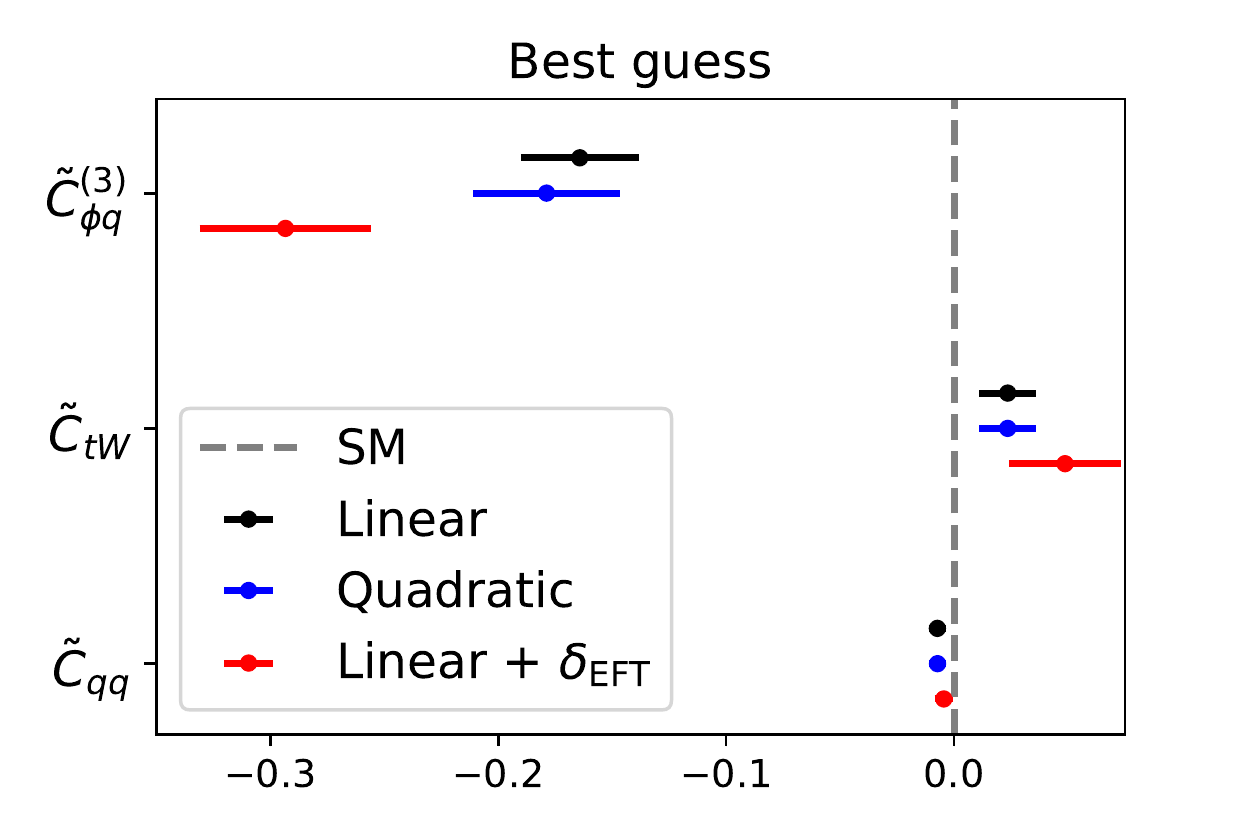}
		\end{center}
		\caption{Same as Fig.~\ref{Fig:Constraints1D}, but with statistical uncertainties of the data in Tab.~\ref{Tab:Data} scaled to $300\,\text{fb}^{-1}$, assuming present central values, systematic uncertainties and theory uncertainties.}
		\label{Fig:300_1D}
	\end{figure}
	\begin{table*}[]
		\centering
		\caption{Marginalized smallest $\SI{95}{\percent}$ intervals obtained in fits in the 'no correlation' scenario Eq.~\eqref{eq:nc} and the 'best guess' scenario Eq.~\eqref{Eq:bestguess} to the data in Tab.~\ref{Tab:Data}  scaled to $300\,\text{fb}^{-1}$ shown in Fig.~\ref{Fig:300_1D}. The central value is in the center of these intervals.}		
		\label{Tab:future1D}
		\begin{ruledtabular}
			\begin{tabular}{cccc}
				Operators& Linear  & Linear$+\delta_\text{EFT}$  & Quadratic \\\hline
				'No correlation'	&&&\\\hline
				$\tilde C^{(3)}_{\phi q}$  & $[-0.043,\,0.009]$ & $[-0.100,\,-0.012]$  &  $[-0.044,\,0.009]$  \\
				$\tilde C_{tW}$   & $[-0.009,\,0.002]$ & $[-0.028,\,0.010]$ & $[-0.009,\,0.002]$\\
				$\tilde C_{q q}$  & $[-0.012,\,0.005]$ & $[-0.012,\,0.007]$  & $[-0.012,\,0.005]$\\\hline
				'Best guess'	&&&\\\hline
				$\tilde C^{(3)}_{\phi q}$  & $[-0.190,\,-0.138]$ & $[-0.331,\,-0.256]$  &  $[-0.211,\,-0.148]$  \\
				$\tilde C_{tW}$   & $[0.011,\,0.036]$ & $[0.024,\,0.074]$ & $[0.011,\,0.036]$\\
				$\tilde C_{q q}$  & $[-0.011,\,-0.004]$ & $[-0.009,\,0.000]$  & $[-0.011,\,0.004]$\\
			\end{tabular}
		\end{ruledtabular}
	\end{table*}
	\begin{figure*}[ht]
		\begin{center}
			\includegraphics[width=0.49\textwidth]{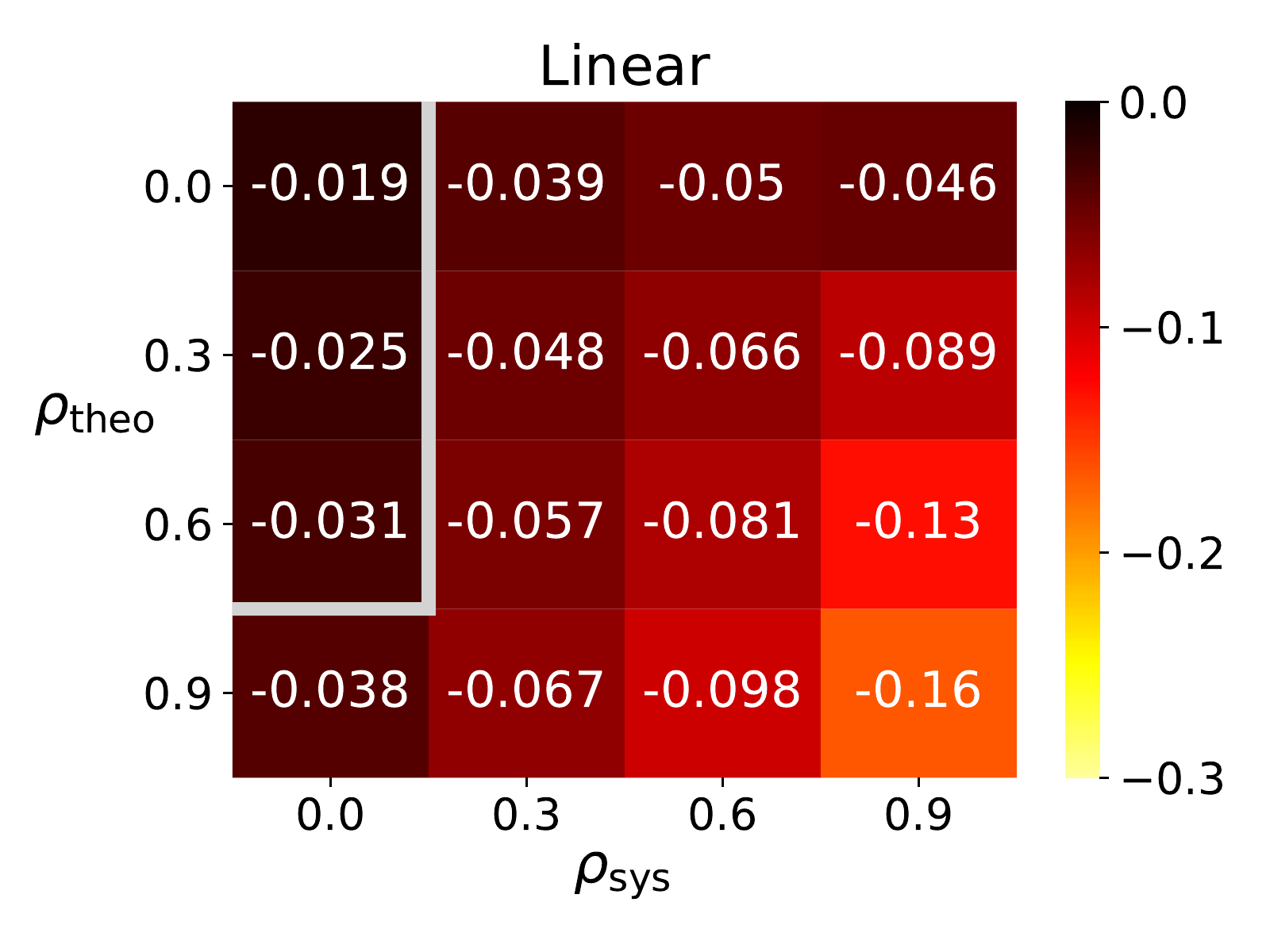}
			\includegraphics[width=0.49\textwidth]{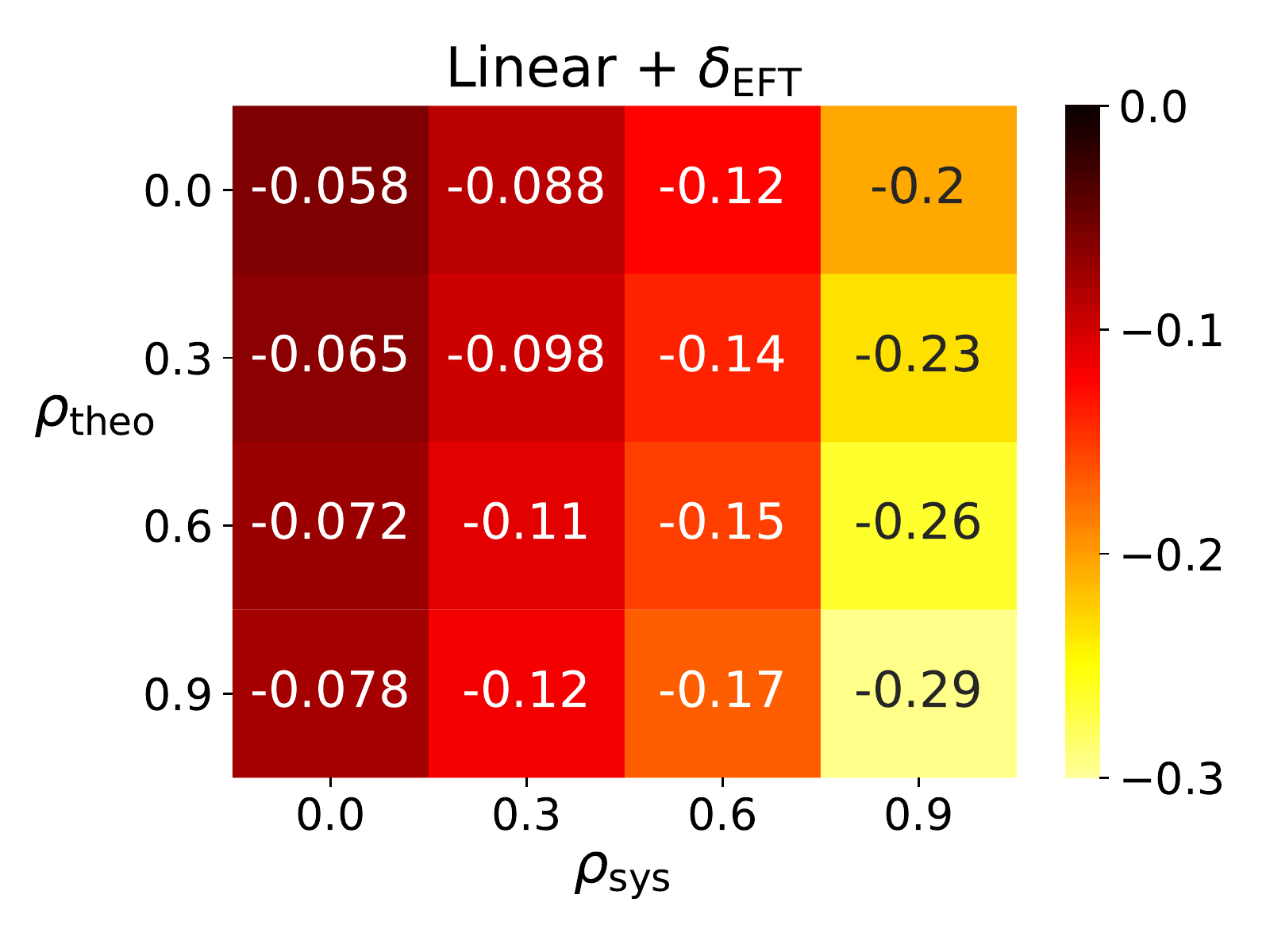}
			\includegraphics[width=0.49\textwidth]{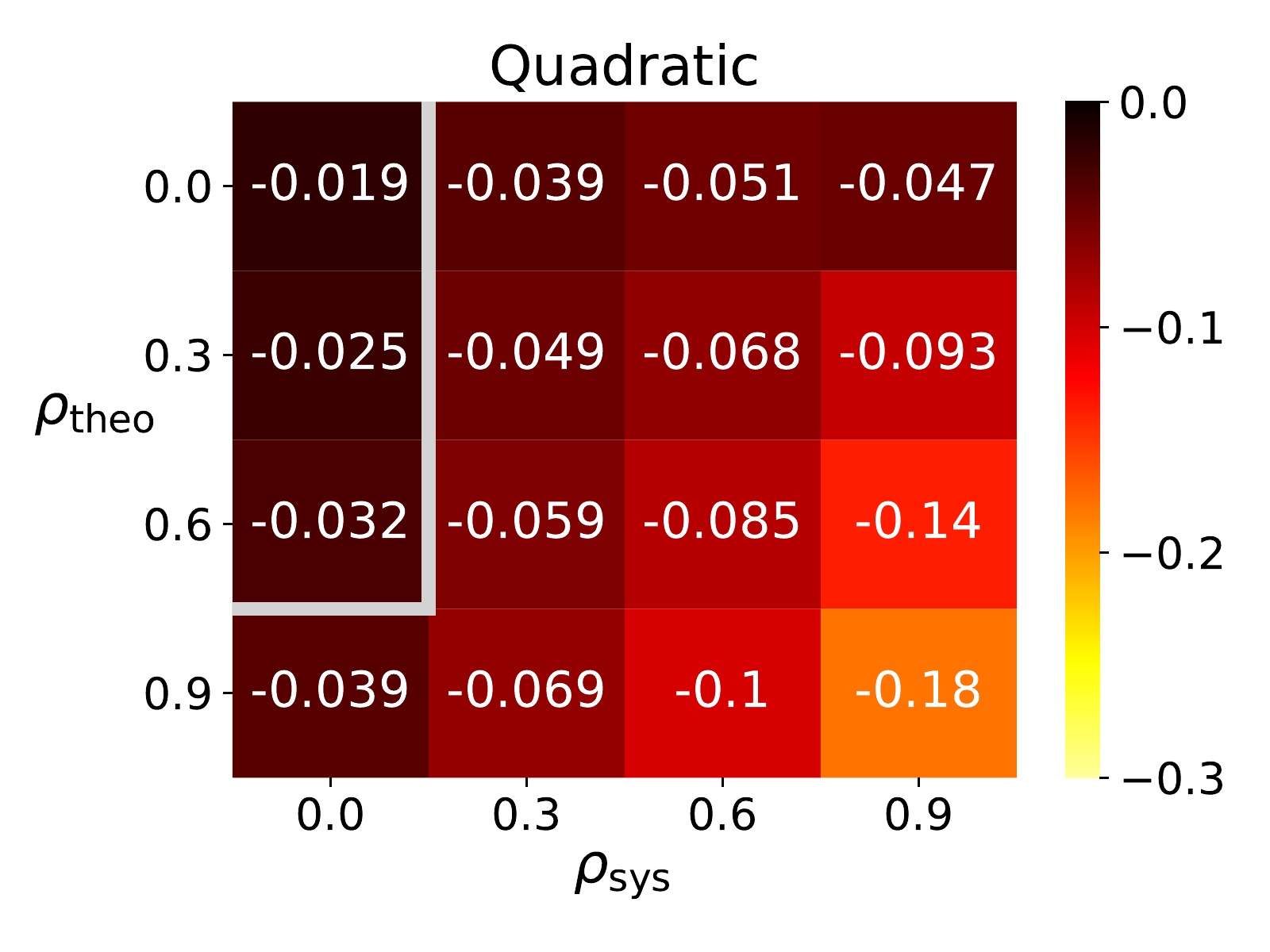}
		\end{center}
		\caption{Same as Fig.~\ref{Fig:Correlation2D_current}, but with all statistical uncertainties of the data in Tab.~\ref{Tab:Data} scaled to the expected integrated luminosity of $300\,\textmd{fb}^{-1}$, 
			assuming present central values, systematic uncertainties and theory uncertainties.}
		\label{Fig:Correlation2DFuture}
	\end{figure*}
	In Fig.~\ref{Fig:300_1D} and Tab.~\ref{Tab:future1D} we give marginal constraints for the coefficients $\tilde{C}_i$ obtained in the 'no correlation' and 'best guess' scenario from fits to data in Tab.~\ref{Tab:Data} with statistical uncertainties scaled to $300\,\text{fb}^{-1}$. 
	Dots and lines denote the central value and the smallest $95\,\%$ interval, respectively.
	We find that increasing the luminosity from up to $20\,\textmd{fb}^{-1}$ for the data in Tab.~\ref{Tab:Data} to $300\,\textmd{fb}^{-1}$ improves the constraints on the coefficients in both correlation scenarios. 

	In contrast, increasing the luminosity further to $3000\,\textmd{fb}^{-1}$ barely improves the constraints (not shown) due to the dominating systematic and theory uncertainties: In the 'no correlation' scenario
	results for $300\,\textmd{fb}^{-1}$ and $3000\,\textmd{fb}^{-1}$ are the same up to percent level for all coefficients. In the 'best guess' scenario the results change by up to $5\,\%$ for $\tilde C^{(3)}_{\phi q}$ and up to $15\,\%$ for $\tilde C_{tW}$ and $\tilde C_{qq}$. 
	The constraints in the light of higher luminosity depend strongly on the correlation scenario:
	In the 'no correlation' scenario the constraints barely change with smaller statistical uncertainties. Comparing the constraints from the data in Tab.~\ref{Tab:Data} to the $300\,\textmd{fb}^{-1}$ projection we find that the central values are slightly shifted and the $95\,\%$ intervals shrink by a factor of 1.2, 2 and 1.3 for $\tilde C^{(3)}_{\phi q}$, $\tilde C_{tW}$ and $\tilde C_{qq}$, respectively. 
	For all coefficients we find agreement with the SM within the $95\,\%$ intervals.

	\begin{table*}[ht]
		\centering
		\caption{Marginalized $\SI{95}{\percent}$ confidence levels from Ref.~\cite{Brivio:2019ius} from a fit to single top-quark total cross sections and top-quark decay data together with the smallest $\SI{95}{\percent}$ intervals obtained in fits in the 'no correlation' scenario Eq.~\eqref{eq:nc} to the data in Tab.~\ref{Tab:Data} excluding differential cross sections. See text for details.}		
		\label{Tab:Comp}
		\begin{ruledtabular}
			\begin{tabular}{ccccc}
				Operators& $\SI{95}{\percent}$ CL \cite{Brivio:2019ius}  & Linear  & Linear$+\delta_\text{EFT}$  & Quadratic \\\hline
				$\tilde C^{(3)}_{\phi q}$ & $[-0.29,\,0.081]$  & $[-0.30,\,0.27]$ & $[-0.30,\,0.28]$  &  $[-0.33,\,0.25]$  \\
				$\tilde C_{tW}$ & $[-0.029,\,0.029]$  & $[-0.013,\,0.013]$ & $[-0.029,\,0.015]$ & $[-0.013,\,0.013]$\\
				$\tilde C_{q q}$ & $[-0.031,\,0.0069]$  & $[-0.115,\,0.095]$ & $[-0.12,\,0.095]$  & $[-0.112,\,0.090]$\\
			\end{tabular}
		\end{ruledtabular}
	\end{table*}
	
	In the 'best guess' scenario the picture changes: We find deviations from the SM for all three coefficients.
	In the case of $\tilde{C}_{tW}$ in the linear and quadratic implementation the $95\,\%$ interval shrinks by a factor of $1.1$ while 
	the central value is shifted to positive values. This leads to deviations of $3.8$ $\sigma$ from the SM. In the linear$+\delta_{EFT}$ implementation the interval grows by a factor of 1.2 while the central value is strongly shifted to larger values leading to a deviation of $4$ $\sigma$. 
	This deviation stems again from differential cross section measurements. The strongly correlated dataset in the 'best guess' scenario prefers larger deviations from the SM. In the combined fit this results in a larger 95~\% interval.
	The central value of $\tilde{C}_{qq}$ grows by a factor of up to $1.5$ in the different scenarios while the $95\,\%$ interval shrinks by a factor of $3.5$, resulting in deviations of up to $5$ $\sigma$ from the SM. 
	Similar to $\tilde C_{tW}$, the origin of the change is due to differential cross section measurements. As correlations deform the 95~\% interval in the three-dimensional parameter space constraints on $\tilde C_{{qq}}$ become stronger while those on $\tilde{C}_{tW}$ and $\tilde C^{(3)}_{\varphi q}$ become weaker.
	In the case of $\tilde C^{(3)}_{\phi q}$ the strongest effects occur. The $95\,\%$ interval reduces by a factor of roughly 1.3, while the central value changes by a factor of up to $1.8$. Again, the deviations origin from differential cross section measurements which push the 95~\% interval further away from the SM as this dataset is most sensitive to the strong correlations. 
	The deviation grow up to $12.6$ $\sigma$ (linear), $11.9$ $\sigma$ (quadratic) and $16$ $\sigma$ (linear$+\delta_\text{EFT}$) so that new physics would be observed. 
	In Fig.~\ref{Fig:Correlation2DFuture} we repeat the analysis from Fig.~\ref{Fig:Correlation2D_current} for the $300\,\textmd{fb}^{-1}$ projection. We find very similar results in all EFT-implementations. 
	Increasing values of $\rho_i$ lead to larger deviations from the SM. 
	Similar to the fit to current data correlations of systematic uncertainties have a stronger impact on the constraints than correlations of theory uncertainties.
	\subsection{Comparison to literature \label{sec:lit}}
	
	As a  consistency check we compare our results to a recent  global SMEFT analysis \cite{Brivio:2019ius}, which
	provides
	$\SI{95}{\percent}$ confidence level intervals from a fit to single top-quark production and top-quark decay data for the coefficients in Eq.~\eqref{eq:WCs}. 
	The dataset used in Ref.~\cite{Brivio:2019ius} is  similar to ours given  in Tab.~\ref{Tab:Data}, except for
	differential cross sections, not taken into account in Ref.~\cite{Brivio:2019ius} and $s$-channel observables, not considered by us.
	To allow for a comparison of results we repeat our fits  for the different 
	fit models, linear, linear$+\delta_{\rm EFT}$, and quadratic, defined in Sec.~\ref{Sec:EFTfitter} in the 'no correlation' scenario  to the data in Tab.~\ref{Tab:Data}, excluding differential cross sections.
	Even though the smallest $\SI{95}{\percent}$ intervals in Bayesian statistics differ from confidence intervals in frequentist statistics, used in Ref.~\cite{Brivio:2019ius}, we expect them to give comparable results. 
	
	In Tab.~\ref{Tab:Comp} we give  the $\SI{95}{\percent}$ confidence levels from Ref.~\cite{Brivio:2019ius}  together with 
	the smallest $\SI{95}{\percent}$ intervals obtained in our fits.
	While the analysis in Ref.~\cite{Brivio:2019ius} utilizes a quadratic ansatz only we show results from all three fit scenarios for comparison as we saw that in the 'no correlation' scenario all three fits give very similar results (see Fig.~\ref{Fig:Constraints1D}).
	Comparing the constraints obtained in our three parametrizations of BSM contributions we find that the results are reasonably similar for all three 	coefficients. While the fit with the linear and quadratic ansatz show excellent agreement, the results for $\tilde C_{tW}$ in the 'linear+$\delta_\text{EFT}$' ansatz show an interval which is larger by roughly 50~\% and shifted to negative values. 
	This is due to the additional EFT uncertainties which affect $\tilde{ C}_{tW}$ stronger as for measurements of $F_{0/L}$ the additional EFT uncertainty can be larger than both the experimental and the SM theory uncertainties(\cite{Khachatryan:2016fky, Aaboud:2016hsq, Khachatryan:2014vma}).
	
	Comparing the results from our quadratic fit (and similarly for the other two scenarios) to Ref.~\cite{Brivio:2019ius} we see some differences but still find reasonable agreement. In the case of $\tilde C^{(3)}_{\varphi q}$ the 95~\% CL is smaller by roughly 35~\% and asymmetric, favoring negative values. As this coefficient simply rescales the SM the additional NLO QCD corrections for SMEFT contributions included in Ref.~\cite{Brivio:2019ius} as well as additional constraints from $tW$, $tZ$ and $s$-channel observables definetly have an impact on the outcome of the fit and are most likely the source of this difference. 
	In the case of $\tilde{C}_{tW}$ constraints from our fit are stronger by a factor of two. This can be traced back to the inclusion of additional $W$ boson helicity fraction measurements from CDF, D0 and CMS. Especially the measurements from Ref.~\cite{Khachatryan:2016fky} give additional constraints on $\tilde C_{tW}$. 
	Constraints on $\tilde C_{qq}$ reported in Ref.~\cite{Brivio:2019ius} are stronger by a factor of 5 compared to our analyses. 
	However, in Ref.~\cite{Brivio:2019ius} an 
	$U(2)_q\times U(2)_u\times U(2)_d$ flavor symmetry is assumed so that contributions from second generation quarks are included. 
	In addition, $s$-channel observables show an enhanced energy dependency compared to $t$-channel observables which we explicitly checked (see also Ref.\cite{Brivio:2019ius}). 
	This provides additional constraints on the four-fermion operator. Togehter with NLO corrections these differences in the dataset and the flavor assumptions can explain the stronger constraints obtained in Ref.~\cite{Brivio:2019ius}.
	
	In summary, differences are expected from the additional NLO QCD corrections to BSM contributions and additional observables which are included in Ref.~\cite{Brivio:2019ius} and from inflated BSM contributions in our linear$+\delta_\text{EFT}$ implementation. 
		
	\section{Conclusions}
	\label{Sec:Summary}
	
	We studied the impact of correlations between (systematic and theory) uncertainties on  multi-experiment, multi-observable  analyses within SMEFT. 
	Specifically, we performed a first quantitative study of such correlations entertaining the example of $t$-channel single top-quark production and top-quark decay.
	This data set allowed us to include 55 measurements from ATLAS, CMS, CDF and DO, given in Tab.~\ref{Tab:Data}, in an analysis to constrain three Wilson coefficients
	(\ref{eq:WCs}).
	We considered different scenarios  for theoretical and systematical uncertainties by varying two parameters $\rho_\text{sys}$ and $\rho_\text{th}$ in the correlation matrices
	based on  simplifying assumptions. 
	We highlighted two scenarios:
	The 'no correlation' scenario Eq.~\eqref{eq:nc}, which has been utilized in previous studies and the 'best guess' scenario Eq.~\eqref{Eq:bestguess}, which incorporates 
	additional correlations between measurements. 
	
	Not unexpectedly, correlations change the constraints on the Wilson coefficients significantly.
	Without correlations no deviations from the SM are found. In the case of strong correlations the SM prediction is not included anymore in the marginalized
	smallest $95\,\%$ intervals of $\tilde C^{(3)}_{\phi q}$, see Fig.~\ref{Fig:Constraints1D}. These deviations can be up to $4.7$ $\sigma$ in the linear and quadratic EFT implementation and $6.6$ $\sigma$ in the 'linear + $\delta_\text{EFT}$' scenario.
 The linear and quadratic models are in very good agreement with each other,  indicating subleading impact of partial higher order EFT corrections only.
	On the other hand,  the 'linear + $\delta_\text{EFT}$' model deviates from the other two approaches, especially in the case of $\tilde{C}^{(3)}_{\phi q}$, and appears to overshoot  the  EFT uncertainty.

	Correlations become even more crucial for future high-luminosity  experiments where the importance of systematic and theory uncertainties is amplified. Assuming  central values fixed, 
	the SMEFT-fit leads to
	significant deviations from the SM at $12.6$ $\sigma$ (linear), $11.9$ $\sigma$ (quadratic) and $16$ $\sigma$ (linear$+\delta_\text{EFT}$) in $\tilde C^{(3)}_{\phi q}$ in the 'best guess' scenario at 300 ${\rm fb}^{-1}$,
	see Fig.~\ref{Fig:300_1D}.
	To benefit from improvements in statistics beyond 300 ${\rm fb}^{-1}$ requires improvements in experimental systematics and theoretical predictions.
	
	Our analysis highlights the importance of correlations in global fits, especially for high-luminosity experiments.
	We suggest to consider different correlation scenarios and to take the corresponding variation into account when presenting results of global fits. 
	At the same time, studies along the lines of Ref.~\cite{Aaboud:2019pkc} are encouraged to provide SMEFT-analyses with the requisite information about 
	correlations of systematic and theory uncertainties. To conclude:
	
	"Correlating uncertainties in global analyses within SMEFT matters in the future even more."

	\begin{acknowledgments}
		C.G. is supported by the doctoral scholarship program of the “Studienstiftung des Deutschen Volkes”.
	\end{acknowledgments}

\end{document}